\documentclass[twocolumn]{aastex62}

\received{2019 07 01}
\revised{2019 09 16}
\accepted{2019 09 22}
\submitjournal{AJ}

\shorttitle{ESO-H$\alpha$ 99 EXor}
\shortauthors{Hodapp et al.}

\begin{document}
\turnoffedit
\title{The New EXor Outburst of ESO-H$\alpha$~99 observed by Gaia ATLAS and TESS}

\correspondingauthor{Klaus Hodapp}
\email{hodapp@ifa.hawaii.edu}

\author[0000-0003-0786-2140]{Klaus W. Hodapp}
\affil{University of Hawaii, Institute for Astronomy, 640 N. Aohoku Place, Hilo, HI 96720, USA}

\author{Bo Reipurth}
\affil{University of Hawaii, Institute for Astronomy, 640 N. Aohoku Place, Hilo, HI 96720, USA}

\author{Bertil Pettersson}
\affil{Department of Physics and Astronomy, Uppsala University, Box 516, SE-751 20 Uppsala, Sweden}

\author{John Tonry}
\affil{University of Hawaii, Institute for Astronomy, 2680 Woodlawn Drive, Honolulu, HI 96822, USA}

\author{Larry Denneau}
\affil{University of Hawaii, Institute for Astronomy, 2680 Woodlawn Drive, Honolulu, HI 96822, USA}

\author{Patrick J. Vallely}
\affil{Department of Astronomy, The Ohio State University, 140 West 18th Avenue, Columbus Ohio, 43210-1173, USA}

\author{Benjamin J. Shappee}
\affil{University of Hawaii, Institute for Astronomy, 2680 Woodlawn Drive, Honolulu, HI 96822, USA}

\author{James D. Armstrong}
\affil{University of Hawaii, Institute for Astronomy, 34 ‘Ohi‘a Ku St., Pukalani, HI 96768, USA}

\author{Michael S. Connelley}
\affil{University of Hawaii, Institute for Astronomy, 640 N. Aohoku Place, Hilo, HI 96720, USA}

\author{C. S. Kochanek}
\affil{Department of Astronomy, The Ohio State University, 140 West 18th Avenue, Columbus Ohio, 43210-1173, USA}

\author{Michael Fausnaugh}
\affil{MIT Kavli Institute for Space and Astrophysics Research, 77 Massachusetts Avenue, 37-241, Cambridge, MA 02139, USA}

\author{Rolf Chini}
\affiliation{Astronomisches Institut, Ruhr-Universit{\"a}t Bochum, Universit{\"a}tsstra{\ss}e 150, 44801 Bochum, Germany}
\affiliation{Instituto de Astronomia, Universidad Catolica del Norte, Avenida Angamos 0610, Antofagasta, Chile }

\author{Martin Haas}
\affiliation{Astronomisches Institut, Ruhr-Universit{\"a}t Bochum, Universit{\"a}tsstra{\ss}e 150, 44801 Bochum, Germany}

\author{Catalina Sobrino Figaredo}
\affiliation{ Astronomisches Institut, Ruhr-Universit{\"a}t Bochum, Universit{\"a}tsstra{\ss}e 150, 44801 Bochum, Germany }



\begin{abstract}

We report photometry and spectroscopy of the outburst of the young stellar object
\object{ESO-H$\alpha$ 99}. The outburst was first noticed in Gaia alert Gaia18dvc
and later by ATLAS. We have established the outburst light curve
with archival ATLAS ``Orange'' filter photometry, Gaia data,
new $V$-band photometry,
and $J$, $H$, and $K_s$ photometry from IRIS and UKIRT. The brightness has fluctuated 
several times near the light curve maximum. The TESS satellite observed ESO-H$\alpha$~99
with high cadence during one of these minor minima and found brightness fluctuations
on timescales of days and hours. Imaging with UKIRT shows the outline of an outflow
cavity, and we find one knot of $H_2~1-0~S(1)$ emission, now named \object{MHO 1520}, on the symmetry axis of
this nebula, indicating recent collimated outflow activity from ESO-H$\alpha$~99.
Its pre-outburst SED shows a flat FIR spectrum, confirming
its early evolutionary state and its similarity to other deeply embedded objects \edit1{in the broader EXor class}.
\edit2{
The pre-outburst luminosity is 34 {$L_\odot$}, a much higher luminosity than typical EXors,
indicating that ESO-H$\alpha$ 99 may be a star of intermediate mass.
}
Infrared and optical spectroscopy show a rich emission line spectrum, including \ion{H}{1} lines,
strong red \ion{Ca}{2} emission, as well as infrared CO bandhead emission, all characteristic
EXors in the broadest sense. Comparison of the present spectra with an optical
spectrum obtained in 1993, presumably in the quiescent state of the object, shows
that during the present outburst the continuum component of the spectrum has
increased notably more than the emission lines. 
The H$\alpha$ equivalent width during the outburst is down to one half of its 1993 level,
and shock-excited emission lines are much less prominent.

\end{abstract}



\keywords{
infrared: stars ---
stars: formation ---
stars: protostars ---
stars: variables: other ---
ISM: jets and outflows
}


\section{Introduction} \label{sec:intro}

Young Stellar Objects (YSOs) 
in Spectral Energy Distribution (SED) classes I and II, 
i.e. stars in their late accretion phase,
often show substantial
variability due to instabilities in the accretion process.
The accretion characteristics of young stars have recently
been reviewed by \citet{Hartmann.2016.ARA&A.54.135H} and we
follow their general line of discussion and the references therein.

Traditionally, the photometric outbursts caused by
increased accretion rates were divided by
\citet{Herbig.1977.ApJ.217.693H} into two classes: 
FU Orionis objects (FUor) and EX Lupi objects (EXor).
The outburst amplitude is similar for both classes, but the
FUor outbursts last for decades to centuries, while
EXor outbursts last from months to maybe a few years.

The first known outburst of a young stellar object, FU Orionis, still remains 
the most substantial of these accretion instability events, having
hardly declined in brightness from its maximum as first discussed by
\citet{Herbig.1977.ApJ.217.693H}. 
For a recent comparison of FUor-type light curves see \citet{Hillenbrand.2019.ApJ.874.82H},
who compare the recently discovered FUor PTF14jg with 
the classical examples, and the comprehensive review of eruptive YSOs by
\citet{Audard.2014.prpl.conf.387A}.
EXors, on the other hand, are typically repetitive on timescales
of a few years to decades, as illustrated in the case of the
deeply embedded EXor V1647 Ori in a series of papers by
\citet{Reipurth.2004.ApJ.606L.119R},
\citet{Aspin.2006.AJ.132.1298A}, 
\citet{Aspin.2008.AJ.135.423A},
\citet{Aspin.2009.AJ.138.1137A},
\citet{Aspin.2009.ApJ.692L.67A},
\citet{Aspin.2009.AJ.137.2968A},
and \citet{Aspin.2011.AJ.142.135A},
and more broadly reviewed by \citet{Audard.2014.prpl.conf.387A}.

While most EXors have shown repeated outbursts on timescales of 
many years to decades, repetitive accretion instabilities on timescales
from weeks down to hours have been found 
\edit1{in many other YSOs},
e.g. the YSOVAR studies of NGC~2264
by \citet{Cody.2014.AJ.147.82C} and \citet{Stauffer.2014.AJ.147.83S}
and the Kepler-2 variability study by \citet{Cody.2017.ApJ.836.41C}.
In many young stars still surrounded by substantial disks, both
accretion instabilities (outbursts) and extinction variations
(dipper events) are observed, and in some cases it can be difficult
to discern which of the two mechanisms dominates.

The two classical types of YSO outburst are also distinct spectroscopically.
EXors show a rich emission line
spectrum probably produced in optically thin funnel flows in
a magnetospheric accretion scenario and veiling of photospheric
absorption lines by an UV and optical continuum produced in
high-temperature shocks. The more
substantial FUor outbursts show a low-gravity absorption line
spectrum reminiscent of a supergiant photosphere thought to be
caused by a self-luminous optically thick accretion disk in a
scenario where the accretion rate from that disk has overwhelmed
the stellar magnetosphere and pushed it back to the stellar surface
\citep{Hartmann.2016.ARA&A.54.135H}.
However, FUor spectra are distinct from those of supergiant photospheres with
their characteristic shift of the spectral type with
wavelength. Spectra in the near-infrared
are classified as a later (cooler) spectral type than the optical spectra.

\edit1{
The original prototypical objects FU Orionis and EX Lupi
used by 
\citet{Herbig.1977.ApJ.217.693H} 
were discovered using photographic methods best suited for
bright and blue objects.
}
As more and more YSO outbursts have been observed thanks to
better all-sky monitoring 
\edit1{and infrared surveys}
these newer objects begin to fill a continuum of 
light curve characteristics such as amplitude, rise time, and rate of decline.
Also, some 
YSO outbursts defy classification
into the FUor or EXor types, either because their outburst duration or spectrum falls 
between that of the classical classes, or because the spectrum
defies classification. 
Examples of such outbursts with intermediate characteristics between
classical FUor and EXor outbursts are V1647 Ori 
\citep[and references therein]{Aspin.2011.AJ.142.135A},
ASASSN-13db \citep{SiciliaAguilar.2017.AA.607A.127S.ASASSN13db}
and PV~Cep \citep{Lorenzetti.2011.ApJ.732.69L, Lorenzetti.2015.ATel.7935.1L, Kun.2011.MNRAS.413.2689K}.
\edit1{
Such outbursts with intermediate characteristics, in particular
more deeply embedded than EX Lupi itself, being associated
with reflection nebulosity and outflow features, and being more
luminous than EX Lupi, have been 
called ``Newest EXors'' by
\citet{Lorenzetti.2012.ApJ.749.188} and
``MNors'' by \citet{ContrerasPena.2017.MNRAS.465.3011C, ContrerasPena.2017.MNRAS.465.3039C}
However, even more extreme transition objects exist
that call into question whether we know enough about the
properties of young eruptive variables to properly define
new classes of this phenomenon.
}

For example, the deeply embedded YSO SVS~13 = V512 Per that shows an emission-line spectrum
was 
\edit1{originally}
listed as an EXor \citep{Aspin.1994.A&A.288.803A,
Eisloeffel.1991.ApJ.383L.19E}.
However, the light curve differs from
that of typical EXors,
since it has not returned
to the pre-outburst brightness 25 years after the outburst
\citep{Hodapp.2014.ApJ.794.169H}.
As an extreme case one outburst with a duration of decades and with a completely line-free
dust continuum spectrum, OO~Ser, was observed by
\citet{1996ApJ...468..861H,2012ApJ...744...56H} 
\edit1{and originally labelled a ``deeply
embedded outburst star (DEOS)''.}

The total number of known YSO outburst objects is still quite small,
at most a few dozen objects of the FUor, EXor, and intermediate cases combined,
even though the discovery rate has been improving in recent years thanks
to several all-sky \edit1{and infrared} monitoring projects.
Every newly discovered object therefore deserves careful analysis.
We report here on the preliminary characterization of ESO-H$\alpha$ 99,
\edit2{
also listed as \object{IRAS08370$-$4030},
}
located at 
08:38:55.17 -40:41:17.34 (J2000.0) 
in the Vela Molecular Ridge,
whose outburst was first noted by Gaia alert Gaia18dvz on 2018 Dec. 19
\footnote[1] {http://gsaweb.ast.cam.ac.uk/alerts/alert/Gaia18dvz/}.

\section{ESO-H$\alpha$~99: Context and Pre-outburst Properties}

\subsection{Context}

The Vela Molecular Ridge (VMR) is a massive molecular cloud complex
straddling the border of Vela and Puppis along the Galactic plane
between Galactic longitudes 259$^\circ$ and 272$^\circ$. It was first
noted by \citet{Dame.1987.ApJ.322.706} and \citet{May.1988.AAS.73.51}. 
\citet{Murphy.1991.AA.247.202}
identified four separate clouds, which they called A, B, C, and
D. A detailed CO survey of the entire VMR was presented by 
\citet{Yamaguchi.1999.APSJ.51.775}.
The distance to the VMR is poorly known, over the
years numerous estimates have been made to individual regions in the
VMR, ranging from 0.7~kpc to 2~kpc
\citep{Pettersson.2008}. 
\edit1{
The Gaia-DR2 parallax of 
ESO-H$\alpha$~99
\edit2{
is ``1.2518 $\pm$ 0.3176 mas'' ($\approx$ 800 $\pm$ 300 pc),
}
not a very precise value due to the faintness of this object.}
Here we adopt a newly determined
distance to VMR-D of 1007$\pm$30~pc
determined by Pettersson and Reipurth (in preparation) based on Gaia-DR2
\edit1{
parallax data of all H$\alpha$ emitting stars in this region.}
This new distance measurement places the
VMR between the foreground Gum Nebula and the background
Carina-Sagittarius arm
\citep{Pettersson.2008}, producing a complex line-of-sight. In
particular it should be noted that the VMR is seen behind the OB 
association Trumpler~10. Star formation is abundant all along the VMR 
as was pointed out by 
\citet{Liseau.1992.AA.265.577},
\citet{Giannini.2007.ApJ.671.470}, and
\citet{Massi.2007.AA.466.1013}.

\begin{figure}[h]
\begin{center}
\includegraphics[angle=0.,scale=0.45]{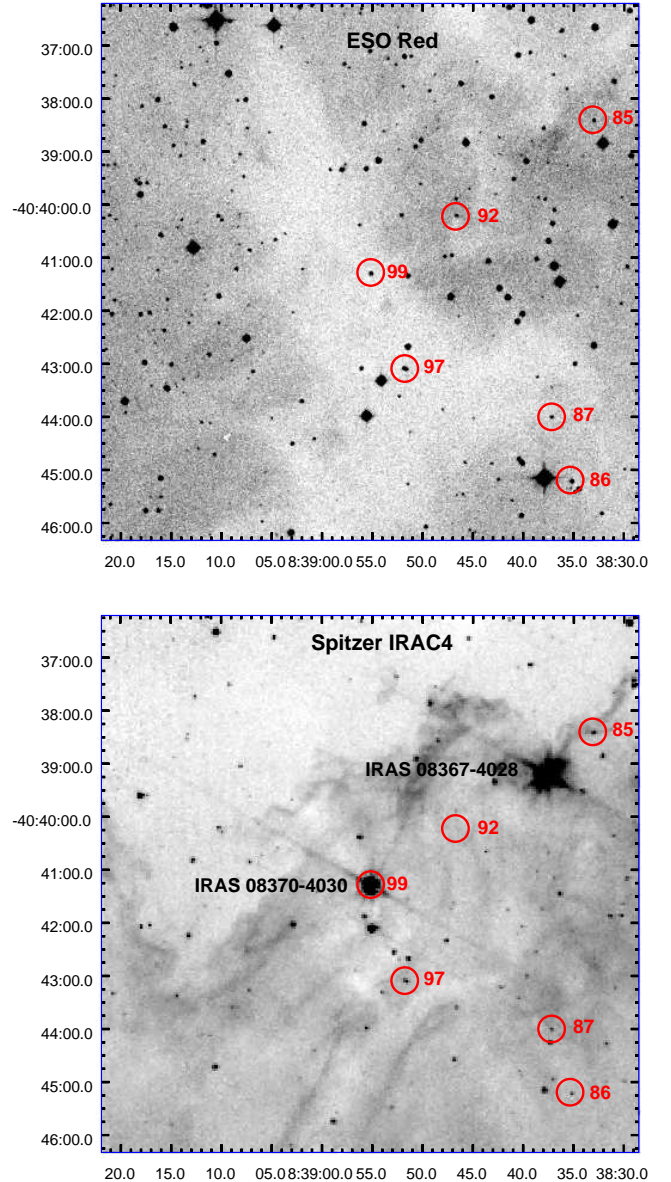}
\caption{
Overview of the region containing ESO-H$\alpha$~99 at optical (ESO Red) and infrared (Spitzer IRAC4) wavelengths.
The features of the foreground extinction visible in the red image is uncorrelated to the PAH emission dominating
the IRAC4 image, illustrating the complicated line of sight. ESO-H$\alpha$ stars are marked and numbered.
ESO-H$\alpha$~99 and
IRAS08370$-$4030 are the same object, with the caveat of possible beam contamination in the IRAS data.
}
\end{center}
\end{figure}

The northwestern part of the ridge is VMR-D, which is dominated by
three HII regions, RCW~27, 32, and 33 
\citep{Rodgers.1960.MNRAS.121.103}.
All of these regions are actively forming stars, which have
been studied at multiple wavelengths, e.g., by
\citet[and references therein]{Strafella.2010.ApJ.719.9,
Strafella.2015.ApJ.798.A104}.
It appears that RCW~27 is a particularly fertile star-forming region.
\citet{Pettersson.1994.AAS.104.233}
carried out a large survey for H$\alpha$ emission stars towards
RCW~27, 32, and 33, many of which have subsequently been identified as
T~Tauri stars, e.g., by 
\citet{Prizinzano.2018.AA.617.A63}. 
One of these young
emission-line stars is ESO-H$\alpha$~99, the object of this paper. 
ESO-H$\alpha$~99 was detected by IRAS and is the dominant flux source
of IRAS~08370-4030.
This star is partly embedded in the dense high-extinction cloud
Sandqvist~1, which is seen against the RCW~27 HII region 
\citep{Sandqvist.1976.AA.53.179}. 
Prominent bright rims are seen in the infrared Spitzer image (Figure 1, bottom panel),
which are likely excited by the OB star in RCW~27. In contrast,
no bright rims are seen in the red optical image (Figure 1 top panel)
suggesting that we are seeing the Sandquist~1 cloud from the un-illuminated
back side.

\subsection{Pre-outburst SED}

Figure~2 shows the spectral energy distribution (SED) of ESO-H$\alpha$~99
based on ground-based and space-based survey data obtained from
VizieR.
The infrared and far-infrared
data are from the compilation of catalog data by \citet[and references therein]{Abrahamyan.2015.AC.10.99A}
and include data from WISE, MSX, AKARI, and IRAS.
The 2MASS data are from the point source catalog \citep{Skrutskie2006}.
The photographic data points (DSS) are from the guide-star catalog of
\citet{Lasker.2008.AJ.136.735L}.
Gaia data are from data release DR~1 
and DR~2
\citep{Gaia.2016.A&A.595A.2G,Gaia.2018.A&A.616A.1G}.
Other optical data are from \citet[and references therein]{Prizinzano.2018.AA.617.A63}.
\edit1{
Note that the SED in Figure~2 does not include the IRAS 100 $\mu$m data point
because of the low spatial resolution of that data point and likely possible
contamination from neighboring far-infrared sources, the closest bright one
being IRAS08367-4028.
}

\edit1{
The most important unsubstantiated assumption in this SED is that all
the measurements refer to the quiescent state of the object.
We do not have a well measured light curve of this object prior
to the start of regular monitoring by Gaia and ATLAS to place
any of the SED data points into the context of a light curve.
It is quite possible that the substantial scatter in the 
flux measurements at optical wavelengths, well above the 
errors indicated in Figure~2, is due to variability.
}
The SED is a typical ``flat spectrum'' distribution
\citep{Greene.1994ApJ...434..614G}. The flat part of the SED extends from
the near-infrared $K$ band at 2.15 $\mu$m out to 160 $\mu$m, characterizing ESO-H$\alpha$~99
as a YSO in transition from a deeply embedded infrared object (Class I)
to a moderately embedded classical T Tauri star.
\edit1{At optical and near-infrared wavelengths, the SED rises steeply, indicating
substantial dust obscuration in the line of sight.}
The integral pre-outburst luminosity of ESO-H$\alpha$~99 from optical
wavelengths to the AKARI 160 $\mu$m data point, a lower limit to the
bolometric luminosity
\edit1{since we do not have sub-mm data to complete the SED}, 
is 34 {$L_\odot$},
adopting again the distance of
1007 pc (Pettersson \& Reipurth, in preparation). 
\edit1{
This pre-outburst luminosity is high compared to most other
T Tauri stars, which are thought to be the precursors of EXor and FUor 
outbursts, and instead is up in the range of
some Herbig Ae/Be stars, specifically the
``group 2'', the flat-spectrum objects as defined by \citet{Hillenbrand.1992.ApJ.397.613H}.
}

\begin{figure}[h]
\begin{center}
\includegraphics[angle=0.,scale=0.45]{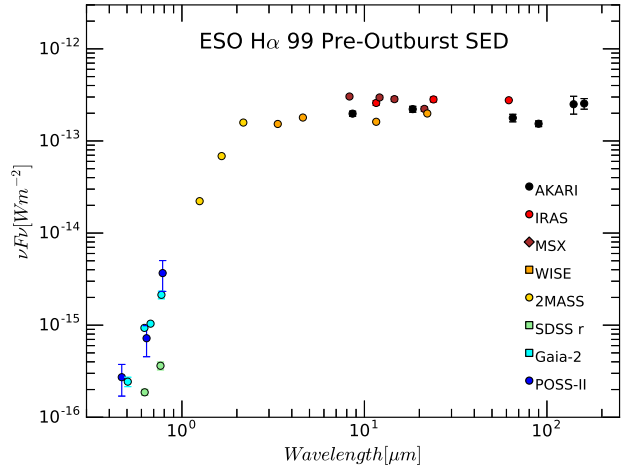}
\caption{
Pre-outburst spectral energy distribution (SED) of ESO-H$\alpha$ 99
from catalog data available
in the VizieR data base. Individual references are in the text.
The SED longward of
the $K_s$ band is remarkably flat.
}
\end{center}
\end{figure}

\section{Observations and Results}

We first noted the rise in brightness of ESO-H$\alpha$~99 in data
from the Asteroid Terrestial-Impact Last Alert System (ATLAS) project, and soon realized that this object
had already been noted in Gaia 
\citep{Gaia-2016A&A...595A...1G} alert
Gaia18dvz posted on 2018 Dec. 19 by the 
Photometric Science Alerts Team (http://gsaweb.ast.cam.ac.uk/alerts)
with the description ``Candidate YSO brightens by more than 1 mag''. 
The Gaia unfiltered (G-band) photometry
covers the years from 2015 to 2019, but has a long gap from 2016 April 27
to 2017 Nov. 2.

\subsection{UKIRT Imaging}

\edit1{
Deep infrared images
were obtained in the $J$ and $H$ bands with the Wide-Field Camera (WFCAM) described by \citet{Casali2007} on UKIRT in March
and April of 2019.
The source  reached $K_s\approx$~8 mag near maximum brightness, leading to saturation
in the $K$-band
in the shortest integration times possible with WFCAM on UKIRT.
We used the saturated UKIRT $K$-band images primarily to show the reflection nebula
in Figure 3.
We also used UKIRT for deep imaging in the $1-0~S(1)$ filter to search
for shock-excited outflow features and the one emission knot that we
found is indicated in Figure~3.
}

\begin{figure}[h]
\begin{center}
\includegraphics[angle=0.,scale=0.45]{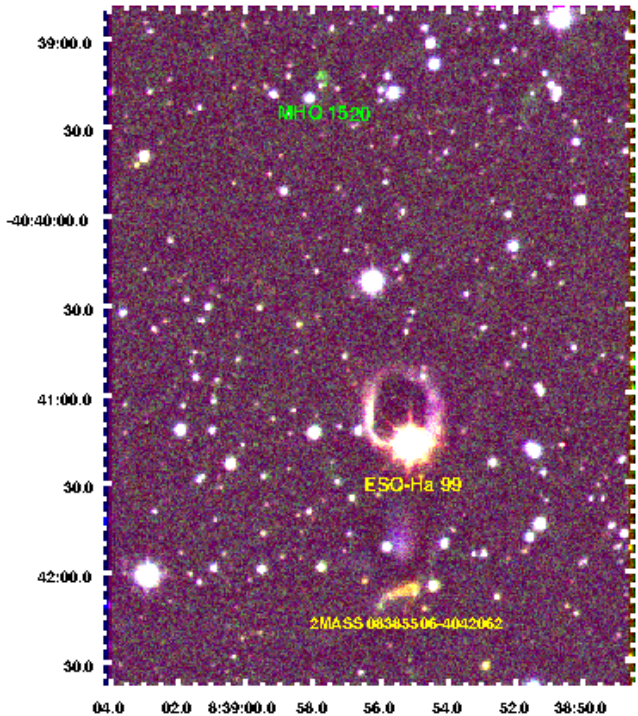}
\caption{
UKIRT color composite of $H$, $S(1)$ and $K$-band images of ESO~H$\alpha$~99
}
\end{center}
\end{figure}

Figure~3 shows a $H$, $S(1)$, and $K$ RGB color composite image obtained with UKIRT. The star ESO-H$\alpha$~99
is associated with a reflection nebula that outlines what appears to be the walls of an outflow cavity.
Farther to the north, 72\arcsec~from the star at P.A. 14\arcdeg, a single small patch of $S(1)$ emission with vaguely
indicated bow-shock shape, lying roughly on the axis of the
reflection nebula, strongly suggests a shock-excited bow shock in a well-collimated outflow. This feature is clearly seen
in the UKIRT $S(1)$  images and was confirmed by $S(1)$ images with lower resolution from the IRIS telescope. The emission
feature is also present on Digital Sky Survey red plates, but not on B and I plates. An association
of this emission feature with ESO-H$\alpha$~99 cannot be conclusively proven without proper motion data, but is 
strongly suggested based on its morphology and location on the symmetry axis of
the reflection nebula.
This shock front has now been included in the catalog of Molecular Hydrogen Objects \citep{Davis..2010A&A...511A..24D}
as MHO~1520.
\footnote[2] {The MHO catalog is now hosted by D. Froebrich at the University of Kent, U.K.}
\edit2{
About 50$\arcsec$ to the south of ESO-H$\alpha$~99
a separate reflection nebula and some $S(1)$ emission associated with the 2MASS
source 08385506$-$4042062 
are visible in Figure~3.
}
\edit1{
2MASS 08385506$-$4042062 
}
is not visible at optical wavelengths, and is
fainter than ESO-H$\alpha$~99 at all wavelengths shown here, indicating
relatively low luminosity. It contributes to the flux of the far-infrared source IRAS08370$-$4030 but is only
a minor contribution at any wavelength.

\subsection{ATLAS}
Most of the optical light curve of ESO-H$\alpha$~99 is based on archival data from the
ATLAS project described by \citet{Tonry.2018PASP..130f4505T}. 
Photometry in the ATLAS ``Orange'' ($O$) filter is shown in Figures~4 and 5.
ATLAS usually takes more than one image of any given region of the sky in each suitable
night to follow fast-moving asteroids. For the light curves in Figures~4 and 5, 
we have median-combined the individual measurements
for each night when the photometric zero points and sky brightness were stable.

\begin{figure*}[h]
\begin{center}
\includegraphics[angle=0.,scale=0.8]{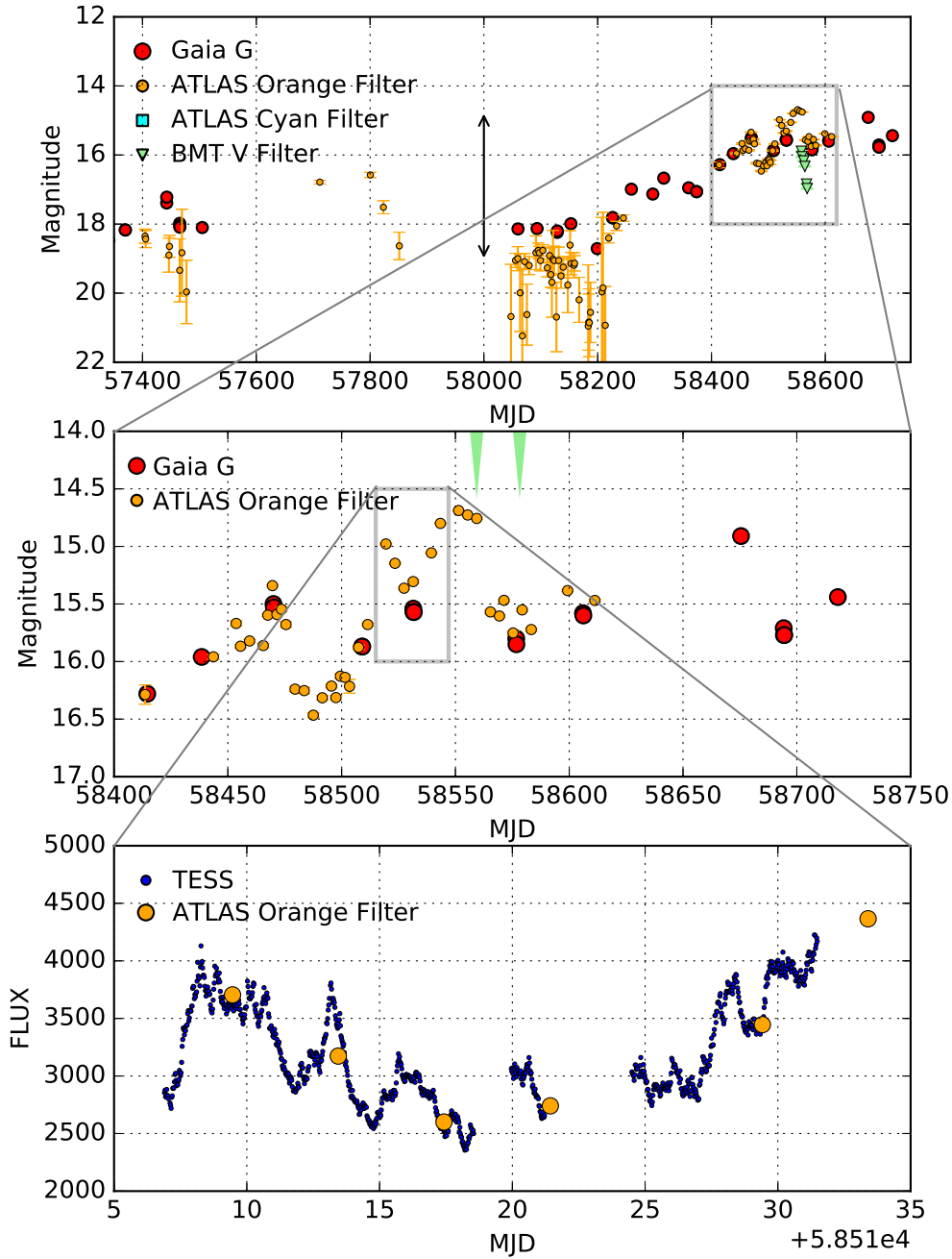}
\caption{
The lightcurve of ESO-H$\alpha$~99
based on archival ATLAS $O$ data, Gaia archival photometry, and TESS photometry.
The top panel shows the full light curve from the beginning of the Gaia mission.
The outburst amplitude in the ATLAS $O$ band from the pre-outburst typical
magnitude ($O$ $\approx$ 19.1 ) to the maximum of $O$ = 14.69 mag is indicated by an arrow.
The middle panel shows the outburst light curve with better resolution.
The green arrows indicate the times and light curve points when the two optical spectra
(Figures 4 and 7) were obtained.
The lower panel shows the TESS photometry calibrated against four coinciding ATLAS data points. 
}
\end{center}
\end{figure*}

\begin{figure*}[h]
\begin{center}
\includegraphics[angle=0.,scale=0.65]{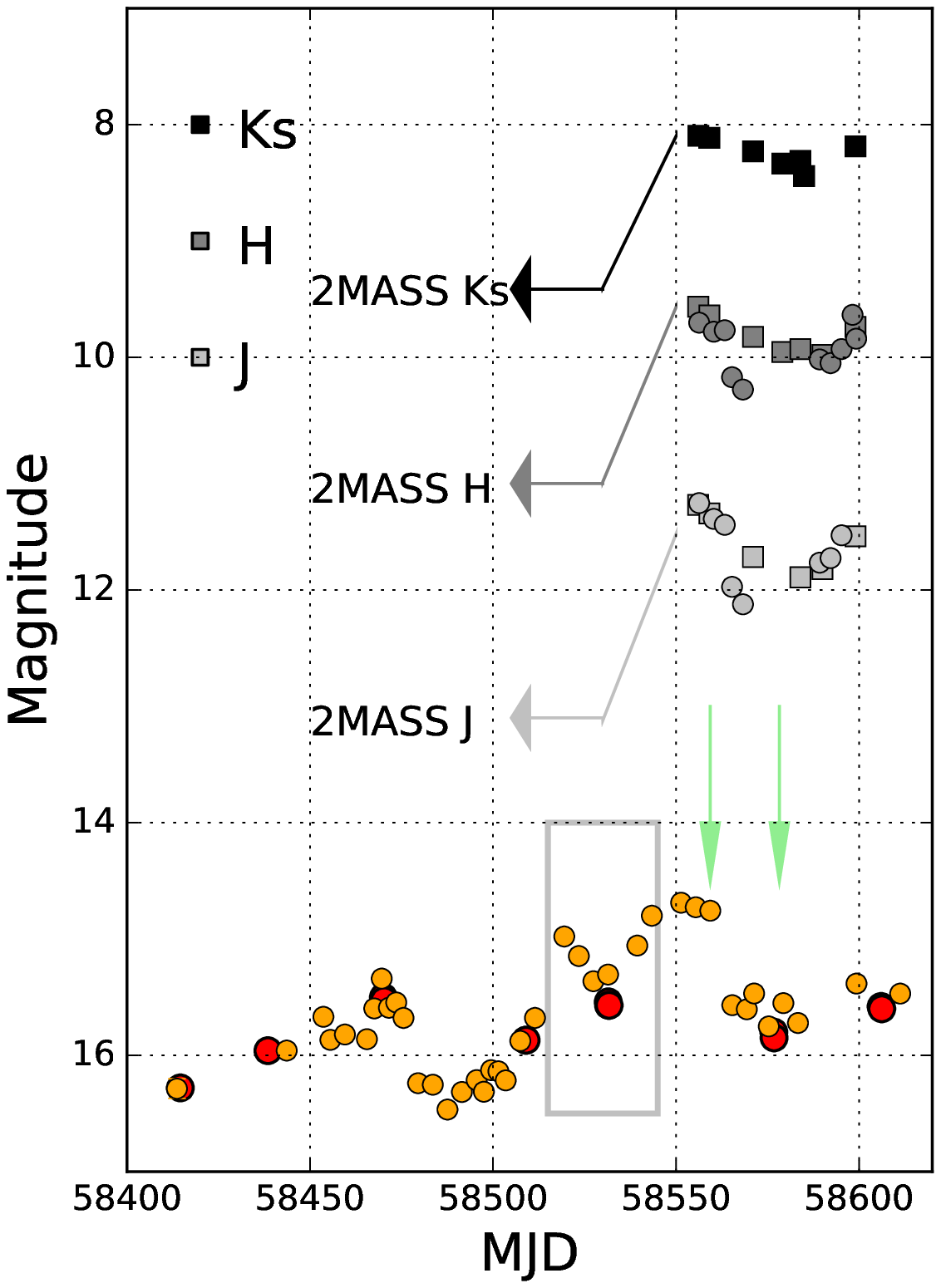}
\caption{
The lightcurve of ESO-H$\alpha$~99 at and after the maximum.
based on archival ATLAS $C$ and $O$ data, Gaia archival photometry,
and post discovery optical and infrared photometry from BMT, IRIS, and UKIRT described
in more detail in the text. 
The left-pointing arrows indicate the pre-outburst
magnitudes from the 2MASS point source catalog.
The green arrows indicate the times and light curve points when the two optical spectra
(Figures 7 and 8) were obtained.
As in Figure 4, orange circles are ATLAS $O$ magnitudes, and red circles are Gaia G magnitudes.
}
\end{center}
\end{figure*}

\subsection{TESS Photometry}

During its Cycle 1 Sector 8 observations, the Transiting Exoplanet Survey Satellite 
(TESS) described by
\citet{Ricker.2015.JATIS.1a4003R}
observed the ESO-H$\alpha$~99 outburst
from 2019-Feb-02 to 2019-Feb-28 with Camera 3. 
We have reduced these observations using an image subtraction pipeline 
optimized for TESS Full-Frame Images. 
This pipeline has previously been applied successfully 
to TESS observations of supernovae 
\citep{Vallely.2019.MNRAS.1385V, Fausnaugh.2019.arXiv190402171F}
and of the tidal disruption event ASASSN-19bt 
\citep{Holoien.2019.arXiv190409293H}.
Because the reference image was constructed from images containing a 
considerable amount of flux from the EXor, 
fluxes in the raw difference light curve are systematically 
lower than their intrinsic values. 
We correct for this discrepancy by scaling and 
shifting the differential flux levels to match four 
ATLAS photometry points obtained concurrently with the TESS observations. 
The TESS photometry, as well as these ATLAS observations, are shown in the lower panel of Figure 4.
It covered the second interim minimum in 
the rise of 
ESO-H$\alpha$~99 to preliminary brightness maximum.

\subsection{IRIS, BMT and UKIRT}

We obtained $J$, $H$, and 
$K_s$ band imaging photometry of 
ESO-H$\alpha$~99
with the IRIS 0.8m telescope
\citep{Hodapp2010}
and 1024$\times$1024 2.5~$\mu$m IRIS infrared camera 
of the Universit\"atssternwarte Bochum
on Cerro Armazones, Chile.
In addition to the monitoring with IRIS, we have also obtained deeper images with better spatial resolution
were obtained in the $J$ and $H$ bands with the Wide-Field Camera (WFCAM) as described above.
The WFCAM filters conform to the ``MKO'' standard described by
\citet{Tokunaga2002} and further characterized in
\citet{Tokunaga..MKO.filters..2005PASP..117..421T}.
Both the UKIRT and IRIS data were calibrated against the 2MASS point 
source catalog
\citep{Skrutskie2006} by the Cambridge Astronomical Survey Unit (CASU)
using the procedures described by
\citet{Hodgkin.2009.MNRAS.394.675H}.
The source  reached $K_s\approx$~8 mag near maximum brightness, leading to saturation
in the $K$-band
in the shortest integration times possible with WFCAM on UKIRT.
In $J$ and $H$, both the UKIRT and IRIS photometry were used, while the
$K_s$-band photometry is only based on IRIS data.
The infrared photometry is shown as light curves in Figure~5 and is the
basis for the color-color diagram in Figure~6

\edit2{
Infrared photometry only started in March of 2019 after the
optical outburst had been noted by ATLAS. The $J$, $H$, and $K_s$
light curve essentially started at the maximum brightness and shows one interim minimum. 
The total outburst amplitude
in the infrared can only be determined in comparison with the
DENIS \citep{Epchtein.1997.Msngr.87.27E} and 2MASS \citep{Skrutskie2006}
catalog values: 
DENIS gives the following magnitudes: 
$I$=15.938, $J$=13.063, $K_s$=9.681 obtained on JD 2450099.72
and $I$=15.810, $J$=13.028, $K_s$=9.581 obtained on JD 2451207.69
while 2MASS gives $J$=13.099, $H$=11.085, and $K_s$=9.414
obtained on JD 2451236.55
The USNOA2.0 red photographic magnitude is 16.40
\citep{Monet.1998.USNOA2.0},
consistent with our pre-outburst brightness
in the ATLAS $O$ filter.
The DENIS and 2MASS near-infrared pre-outburst magnitudes are consistent with
each other, and the two DENIS measurements are consistent over an
interval of 1108 days, so we
assume that all these measurements represent the quiescent
state of ESO-H$\alpha$ 99. In Figure~5 the -- presumably quiescent -- 2MASS catalog
magnitudes, whose filter bandpasses match the IRIS photometry,
are indicated by left-pointing arrows.
}

We have also obtained $V$-band photometry with the 40cm
Bochum Monitoring Telescope (BMT) on Cerro Armazones 
\citep{Ramolla.2013.AN.334.1115R},
where the object remained observable 
longer than from Hawaii. The photometry was calibrated against 
a set of in-field non-variable stars from
the APASS all-sky catalog
\citep{Munari.APASS.2014AJ....148...81M} and is included as green triangle
symbols in Figure~4.

\subsection{IRTF Infrared and Faulkes Telescope Optical Spectroscopy}

Finally, on 2019 March 16, UT, we obtained a near-infrared spectrum of ESO-H$\alpha$~99 using the
NASA Infrared Telescope Facility (IRTF) with the SPEX instrument
\citep{Rayner.2003.PASP.115.362}
in short cross-dispersed (SXD) mode with a 0\farcs3 slit, giving
a spectral resolution of R=2000. 
Low-resolution optical spectra of ESO-H$\alpha$~99 were obtained
with the Faulkes Telescope North on Haleakala, Maui, on 2019 March 17 (UT)
and 2019 April 5 (UT)
using a slit width of 1\farcs6 and 30\arcsec~slit length of
the FLOYDS spectrograph.
This resulted in a FWHM of the [\ion{O}{1}] night sky lines of 14~{\AA} and a spectral
resolution of R = 425 at the wavelength of the \ion{Na}{1} doublet.

The optical and infrared spectra from March 17 and 16 (UTC), respectively,
were merged into Figure~7. 
It turned out that these spectra were obtained close to the maximum brightness
reached by ESO-H$\alpha$~99 during this outburst.

The spectrum \edit2{near maximum brightness} in Figure 7, combining optical data from the Faulkes
telescope and near-infrared data from the IRTF/SPEX, 
\edit2{
shows prominent \ion{H}{1}, \ion{He}{1}, \ion{Ca}{2}, and 
CO bandhead emission, the only noticeable exception being \ion{Na}{1} D line in absorption.
Note, however, that \ion{Na}{1} is in emission in the near-infrared. 
Besides the emission lines, a continuum is present, but
at the spectral resolution of our data, we cannot detect
absorption lines that could be used for spectral classification.
The H$_2$ 1--0 S(1) line is weakly detected in emission at
the position of the star itself, indicating shock-excitation
of H$_2$ near the star.
}

\subsection{The pre-outburst spectrum}

As part of a spectroscopic survey of the brighter stars
of the sample of 
H$\alpha$ emission line stars
in the VMR-D region of 
\citet{Pettersson.1994.AAS.104.233}
a spectrum of ESO-H$\alpha$~99 was obtained
on February 22, 1993 at the ESO 3.6m telescope with the OPTOPUS
multi-object spectrograph and an
exposure time of 2 $\times$ 30 min.
The original data are no longer available, but a plot of the
spectrum is. For the analysis in this paper, we have
digitized this plot.
Although we do not know
the photometric brightness of the star at the time this spectrum
was taken, absent any indication of a prior outburst at that time, 
we assume that the spectrum represents the quiescent state.
\edit1{
The pre-outburst spectrum is the one shown in red in the
comparison of spectra in Figure~8.
}

\section {Discussion}

\edit1{
Our goal is to compare the newly discovered and on-going outburst of
ESO-H$\alpha$~99 
to well-studied cases and to place it into the context of 
the ensemble of eruptive events in young stars.
In our discussion, we will use the term ``EXor'' in the broadest sense describing
a young eruptive variable with an emission-line spectrum.
Our use of that term is not intended to imply a close similarity
of the newly discovered object with specifically the prototype EX Lupi.
}

\subsection{Classification of Outburst}

The observed outburst amplitude is strongly wavelength dependent,
being much larger at optical wavelengths than in the near-infrared.
The optical light curve in the ATLAS $O$ filter of ESO-H$\alpha$~99 (Figure~4)
shows a rise of about 4.4 magnitudes from the quiescent brightness \edit2{($\approx$ 19.1 mag)}
up to the maximum of brightness observed until now \edit2{($O$ = 14.69 mag)}. Just prior to
the rise, around MJD 58185, both the ATLAS and Gaia data indicate
a brief ($\approx$30 day) dip in brightness of about 1.3 mag in ATLAS $O$ \edit2{below the quiescent brightness}. 
The immediately following rise in brightness to the maximum is not
steady, rather, the ATLAS data show three \edit1{interim} minima superposed on
the overall \edit1{rise of the} outburst light curve.
As we will discuss below in the context of the spectroscopy,
at least one of these interim minima in ESO-H$\alpha$~99 is associated with
spectral changes. We can exclude that these are merely caused by
extinction variations.
\edit2{
The few sparse Gaia data obtained after 
ESO-H$\alpha$~99 became observable again in July of 2019 show variations
of about 0.5 mag again, similar to the interim minima observed earlier
with both Gaia and ATLAS data. We conclude that the semi-periodic
fluctuations with typical periods of about one month are continuing now superposed on a fairly stable plateau near
maximum brightness.
}

\edit2{
Quasi-periodicity on the order of one month 
as observed in ESO-H$\alpha$~99 superposed on the outburst
is too long for the
expected rotational period of a still rapidly rotating young
star and the magnetically coupled accretion flows, but must be related to regions of the
rotating disk farther away from the star.
}

\edit1{
Similar, though shorter period fluctuation superposed on the 
years-long outburst have been observed by \citet{SiciliaAguilar.2017.AA.607A.127S.ASASSN13db}
in ASASSN 13db and interpreted as rotational modulation of starspots
that are coming and fading on timescales of months.
}

About 2 years before the 2019 major
outburst, the ATLAS data indicate
a minor maximum around MJD 57800 (2017 Feb. 16) 
The coverage of this prior maximum by ATLAS is poor,
\edit1{
and there are no confirming Gaia measurements.}
We can only 
state that this maximum has lasted for at least a few months, 
and that the amplitude above the quiescent brightness was
about 2.5 mag in ATLAS $O$ filter,
substantially less than the present maximum.
Despite the limitations of the available data, 
\edit1{we note that} 
this smaller prior maximum is
entirely consistent with the characteristics of other 
EXors, as was reviewed in the Section 1, and in particular appears similar
to the sporadic outbursts of the prototypical EX Lupi
\citep{McLaughlin.1946.AJ.52.109M, Herbig.2001.PASP.113.1547H, Herbig.2007.AJ.133.2679H}, where
\citet{2010ApJ...719L..50A} distinguish ``characteristic'' and ``extreme''
outbursts. In this terminology, the present 2019 outburst of ESO-H$\alpha$~99
\edit1{
qualifies
as an ``extreme'' outburst, while the smaller 2017 outburst would be ``characteristic''.}

\subsection{Accretion Instability on Timescales of Days}

The outburst of ESO-H$\alpha$ 99 was fortuitously observed
by TESS. The data come from the Full Frame Images and give
photometry on a 30 min cadence. The resulting finely resolved light curve during
the 28 days of observations is shown in the lower panel of Figure~4.
The TESS observations covered the second of the three minor \edit1{($\approx$ month-long)}
interim minima superposed on the overall rise to maximum brightness.
The TESS light curve shows multiple local maxima and minima
with typical rise and fall time of order of one day.
A period analysis 
using the Lomb-Scargle
algorithm developed by \citet{Lomb1976} and \citet{Scargle1982},
and implemented in the Period Analysis Software (PERANSO) written by T. Vanmunster
did not find 
any strict periodicity of these short term
variations, rather, there is significant power in periods
ranging from 2 to 10 days, after ``whitening'' out the power
associated with the overall minor minimum. These variations on timescales
of several days on top of the
longer term features of the light curve are similar to
the variations during quiescence found 
by \citet{Cody.2017.ApJ.836.41C} using K2 data for YSOs with
less substantial disks and ``bursts`` (rather than major ``outbursts'') of order of tens of percent
in flux relative to the quiescent, low accretion state. 

Variability of FU Ori during its on-going outburst was studied
by \citet{Siwak.2013.MNRAS.432.194S.FU.Ori, Siwak.2018.AA.618A.79S.FU.Ori}
and attributed to a small number of unstable accretion ''tongues'' (referred to as ``funnels'' here)
and their associated hotspots, revolving around the star.

\edit2{
Stable accretion funnels and stationary hotspots would lead to
rotationally modulated periodic variability on timescales of
the rotation of the star, i.e., of order of a few days. This is not what
we observe, however. The short term variations observed by
TESS cannot be described as quasi-perodic. At best, they
indicate that ``every few days, something changes''.
This is best understood if we assume that we observe several
active accretion funnels at any given time and their varying
accretion rates, coupled with some rotational modulation
of this complex scenery.
}

\subsection{Near-Infrared Colors and SED}

Figure~6 shows the near-infrared $J-H/H-K$ color-color diagram
of ESO-H$\alpha$ 99 in both quiescence and outburst.
We included similar objects observed by 
\citet{Lorenzetti.2012.ApJ.749.188} and
V2492 Cyg, studied in detail by
\citet{Aspin.2011.AJ.141..196A} for comparison, as
it is the most reddened EXor known to date.

\begin{figure}[h]
\begin{center}
\includegraphics[angle=0.,scale=0.65]{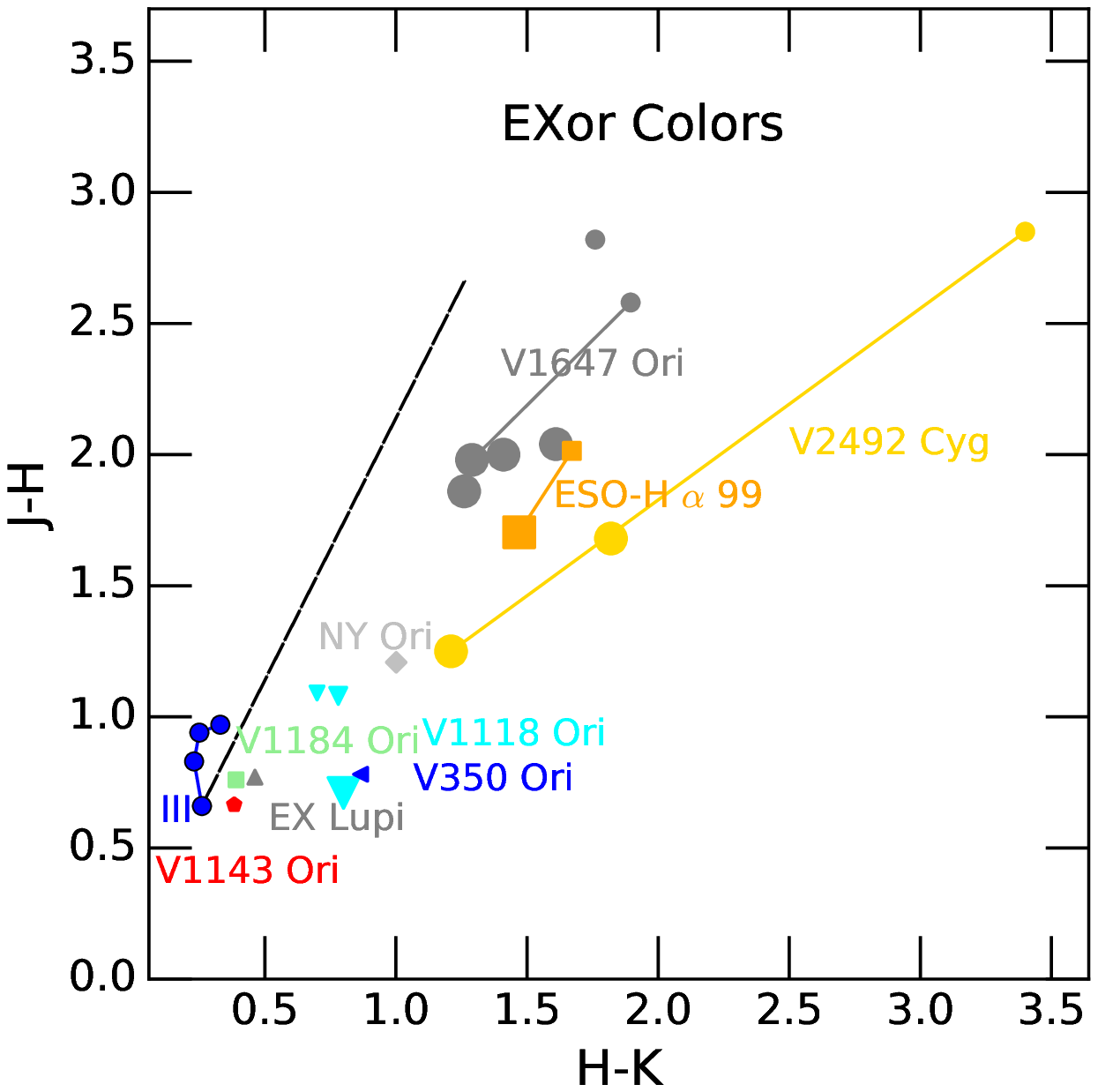}
\caption{
Color-Color diagram ($J-H$ vs. $H-K_s$) of ESO-H$\alpha$~99
in quiescence and outburst (orange symbols). Larger symbols 
represent the bright state of the EXor. For comparison we have
included the same known EXors as
used in \citet{Herbig.2008.AJ.135.637H}.
We have added additional data
on V1647 from \citet[and references therein]{Aspin.2011.AJ.142.135A}
and data on V1118 Ori from
\citet{Giannini.2017.ApJ.839.112}.
The blue filled circles represent the locus of unreddenend class III stars
from \citet{Wegner.2014.AcA.64.261W}
and the dashed line represents the reddening vector due to interstellar
extinction based on the data by
\citet{Straizys.2008.BaltA.17.125}.
}
\end{center}
\end{figure}

Figure 6 shows that ESO-H$\alpha$ 99
shares the colors of the other two deeply embedded EXor objects,
V1647 Ori and V2492 Cyg, and shows similar color changes from
the quiescent state to the outburst maximum.
\edit1{
In general, all EXors get ``bluer'' during outburst and the 
color-color path from quiescent to outburst location is roughly
parallel to the interstellar reddening vector, if anything slightly
less steep. While the color changes are therefore consistent with
a clearing of obscuring material during outburst, this is certainly not
the only mechanism at work in ESO-H$\alpha$~99
since we are also seeing major changes in the line ratios in the spectrum.
Those spectral changes cannot be explained simply by a varying amount
of line-of-sight extinction.
Some contribution from changing extinction, possibly caused by
irregularities in the inner disk orbiting in and out of the line
of sight is quite possible, but cannot be confirmed with the 
available data.
}

We have compared the pre-outburst spectral energy distribution
of ESO-H$\alpha$~99 (Figure~2) with the SEDs of other EXors and
FUors using the VizieR database and its Photometry Viewer.
This database is readily available, so we do not show all SEDs
here. The wavelength coverage of those SEDs is quite heterogeneous,
so we will not discuss the similarities and differences quantitatively.
The SEDs of most
classical, \edit1{i.e., low extinction} EXors, some of which are included in
the near-infrared color-color diagram in
Figure~6 and occupy the low-color-value area of this diagram, have the peak of their SED at near-infrared
wavelengths, and decline from there to the long-wavelength limit
of the available data. 
Only the deeply embedded EXors 
show flat or nearly flat SEDs:
NY Ori is flat from 1 to 20 $\mu$m, but has no data beyond that.
Deeply embedded, very red EXors such as V2492 Cyg, PV Cep, and
V1647 Ori have essentially flat SEDs out to 100 $\mu$m.
The SED of ESO-H$\alpha$~99 is very similar to those three
deeply embedded flat SED EXors. 
This flat 10 - 100 $\mu$m SED also puts
ESO-H$\alpha$~99 in the same SED class as OO~Ser, the
"deeply embedded outburst star" \citep{1996ApJ...468..861H}
that falls outside of the classical FUor vs. EXor classification
scheme.

We also note that the classical FUors FU Ori and V1057 Cyg
have SED peaks at near-infrared wavelengths. 
The FUor V2775 Ori, however,
\citep{Fischer.2012.ApJ.756.99F}
has a flat SED similar to the one discussed here.
Deeply
embedded FUor-like objects, i.e., objects spectroscopically similar to
a FUor but no historically observed outburst, such as L1551 IRS5, have SEDs
rising throughout the 10 - 100 $\mu$m range.

While EX Lupi, the prototypical EXor, is not associated with
much nebulosity and is not deeply embedded, ESO-H$\alpha$ 99
is highly reddened and is associated with a reflection nebula
with a bubble-like morphology, probably light scattered
off the walls of an outflow cavity. Both its red colors and
the presence of a reflection nebula 
support the fact
that ESO-H$\alpha$~99
is a very young object, still deeply embedded in its parent
molecular cloud, and surrounded by a substantial gas and dust
disk.
We have found one
small region of shock-excited $H_2~1-0~S(1)$ emission with indication of a
bow-shock morphology near the symmetry
axis of the reflection nebula, strongly suggesting that
this feature is the result of a highly collimated outflow
from ESO-H$\alpha$ 99. Such highly collimated outflows are
frequently found in YSOs, but are not typical of EXors or FUors. 
We interpret this as additional evidence that
ESO-H$\alpha$~99 is, within the \edit1{broad} EXor class of outbursts,
one of the youngest of this class of objects.

\begin{figure*}[ht!]
\begin{center}
\includegraphics[angle=0.,scale=0.50]{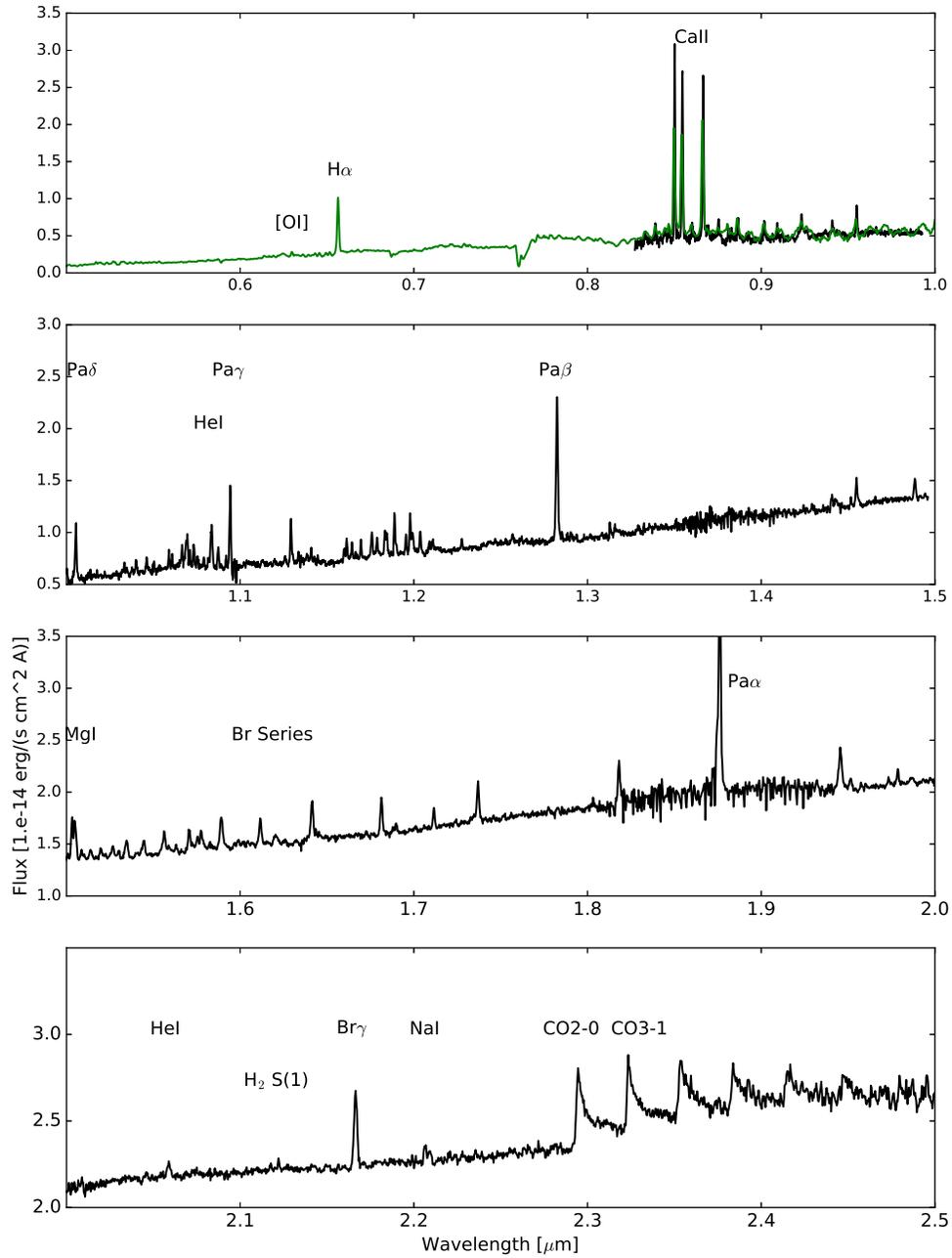}
\caption{
The IRTF/SPEX spectrum from 2019/3/16 UT combined with the optical spectrum (green) taken on 2019/3/17 UT
with the FLOYDS spectrograph on the Faulkes Telescope North on Haleakala.
The optical spectrum has not been corrected for telluric absorption, while the infrared
spectrum has.
The spectrum shows strong emission lines, indicating that the outburst is broadly an EXor type.
}
\end{center}
\end{figure*}

\subsection{Spectral Changes During Outburst}

We compare low-resolution optical spectra of ESO-H$\alpha$~99 at three epochs: 
the pre-outburst spectrum from 1993 Feb. 22, presumably during a quiescent
phase,
the spectrum taken very close to the maximum of the current outburst
on 2019 March 17 \edit2{shown in Figure 7}, and the spectrum taken on 2019 April 5 when the
ATLAS $O$ magnitude was 0.8 mag lower than at the maximum.

\edit1{
We do not have coincident photometry for the pre-outburst
spectrum from 1993 Feb. 22. From Figure 4 it is clear that
ESO-H$\alpha$~99 had shown some variability prior to the
present outburst. On the other hand, none of the pre-outburst
data (POSS, 2MASS etc.) give any indication of a prior large
outburst. Our measured quiescent pre-outburst
brightness in the ATLAS ``Orange'' filter is about $O$ = 19.1 mag
}
\edit2{
We therefore make the 
assumption that at the
time of the 1993 spectrum, the brightness was $O$ $\approx$ 19 mag.
}

\edit1{
For the two epochs in 2019, we have closely coincident ATLAS
``Orange'' filter photometry. For 2019 03 17, ATLAS obtained
photometry in the same night, about 1.25 hours after the spectrum.
For the 2019 04 05 spectrum, the closests ATLAS photometry
is from the following night, 25.7 hours after the spectrum.
}
The epochs of the recent spectra are indicated by blue arrows in
Figures 4 and 5 and place those spectra in context of the light curve.

Of the two 2019 optical spectra, the first was obtained near the maximum of the observed
light curve during this ongoing outburst, and the second in the following minor interim minimum
in the light curve, after which the object brightened again.
\edit1{
In the middle panel of Figure 4 and in Figure 5, the brightness maximum at the time
of the first spectrum and 
the interim minimum on 2019 April 5 are documented by
ATLAS $O$, and infrared $J$, $H$, and $K$ photometry.
}
\edit1{
Table 1 lists the dates, ATLAS $O$ magnitudes, and the equivalent
width of four spectral lines, both permitted and forbidden. 
}

\begin{deluxetable}{lrrr}
\tabletypesize{\scriptsize}
\tablecaption{Optical Spectroscopy}
\tablewidth{0pt}
\tablehead{
\colhead{    } & \colhead{1993 02 22} & \colhead{2019 03 17} & \colhead{2019 04 05}
}
\startdata
Magnitude ($O$) & $\sim$ 19 & 14.76 & 15.55\\
Rel. Flux & 0.02 & 1.00 & 0.48\\
Color in Fig. 8 & red & blue & green\\
$[$\ion{O}{1}$]$ 6300 EW [\AA] & -29.3 &  -2.2 & -4.7 \\
$[$\ion{O}{1}$]$ 6364 EW [\AA]  & -11.5 & low S/N  & -1.4 \\
\ion{H}{1} 6563 EW [\AA]    & -75.1 & -36.5 & -58.9\\
$[$\ion{S}{2}$]$ 6716+6731 EW [\AA] & -19.4 &  -0.9 & -1.8 \\
\ion{Ca}{2} 8498 EW [\AA] & no data & -35.0 & -47.7\\
\ion{Ca}{2} 8542 EW [\AA] & no data & -37.2 & -52.2\\
\ion{Ca}{2} 8662 EW [\AA] & no data & -41.9 & -54.0\\
\enddata
\end{deluxetable}

For now, we can leave the question open whether March 17 marked
the overall maximum on the current outburst, or whether it is
merely a preliminary interim maximum. 
\edit1{After it reemerged from behind the Sun in August 2019, we have obtained a few new Gaia data points
indicating that ESO-H$\alpha$ 99 has experienced another high point near the magnitude
of the March 2019 maximum and is otherwise staying within a magnitude of the maximum
brightness with some variations. 
}
The important point is that changes
in \edit1{broadband photometry} are associated with changes in the \edit1{emission line} spectrum.

The early 1993 optical spectrum of ESO-H$\alpha$~99 
offers some insight into what we assume is the quiescent state of the star:
First, the 26-year
old spectrum shows the forbidden lines of [\ion{O}{1}] at 6300 and 6363~{\AA}~and \edit2{blended} [\ion{S}{2}]
6716/6731~{\AA}, 
which together with strong H$\alpha$ emission are
characteristic of Herbig-Haro shocks. 
The molecular hydrogen shock front MHO~1520 approximately along
the line defined by the outflow cavity associated with
ESO-H$\alpha$~99 (Figure~3) shows that outflow activity must
have occurred in the more distant past, certainly prior to the
present outburst.

Second, a forest of permitted \ion{Fe}{2} and forbidden [\ion{Fe}{2}] lines is
seen in the 5000-5500~{\AA} region. These lines are only seen in the most
active T~Tauri stars 
\citep{Herbig.1962.AdA&A.1.47H}, and the spectrum shows a strong
semblance to that of the HH~32 driving source AS~353A 
\citep{Eisloeffel.1990.A&A.237.369E}
and the HH~46/47 driving source 
\citep{Reipurth.1991.AA.246.511}.

Third, the \ion{Na}{1} doublet at $\lambda$5890/5896~{\AA} is in absorption, 
fairly weak and unresolved
in the 1993 spectrum as well as in the 2019 spectra. 
The \ion{Na}{1} absorption blends with the \ion{He}{1} line at 5876~{\AA}.
Helium emission lines have been observed in several other EXors, e.g.,
NY Ori, V1118 Ori, and V350 Ori by
\citet{Herbig.2008.AJ.135.637H}.

\edit1{
Altogether, the 1993 \edit1{quiescent} spectrum is indicative of a highly active,
strongly accreting YSO star that is driving a shocked
outflow. 
}

\begin{figure}[ht!]
\begin{center}
\includegraphics[angle=0.,scale=0.45]{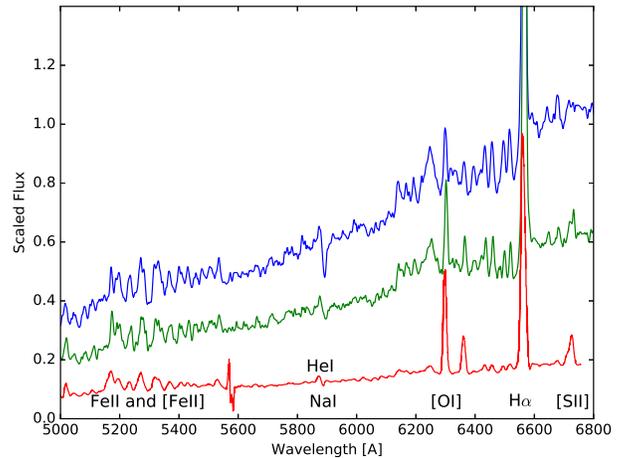}
\caption{
A comparison of \edit2{the optical} spectra at three epochs:
pre-outburst (1993 Feb. 22, red), maximum brightness (2019 March 17, blue), and declining phase (2019 April 5, green).
The spectra are shown qualitatively in the sequence of increasing integrated flux,
but are scaled for clarity and not in proportion to the integrated flux.
Table 1 gives dates, photometric magnitudes and relative flux for the three spectra.
The timing of the two recent spectra relative to the light curve is indicated
by green arrows in Figures 4 and 5.
}
\end{center}
\end{figure}

\edit1{
During the present outburst, the spectrum has changed substantially.
}
Referring to Table I, the equivalent width of 
\edit1{
all the emission lines was much higher in quiescence (1993) than
during the outburst.
}
H$\alpha$ line was much higher
($\sim$75~{\AA}) than in the 2019 March 17 spectrum ($\sim$37~{\AA}). 
On 2019 April 5, \edit1{at a time when the broadband integrated flux was only
about one half of that at maximum, } the spectrum has an H$\alpha$ equivalent width between
those two extreme points: EW $\sim$62~{\AA}.

The spectral changes observed between those two \edit2{2019} spectra clearly
indicate that this interim minimum was not just caused by changing
amounts of obscuring dust in the line of sight, which should not
affect the equivalent width of H$\alpha$. 
Rather, the photometric and spectral variations show real changes
in the components of the accretion flow: inner disk, accretion funnels,
and possibly a surface hot spot on the star.

\edit1{
Comparing the change of H$\alpha$ equivalent width to the
change in broadband flux in Table 1, it is clear that the flux of the
H$\alpha$ line does increase with increasing brightness,
but that the continuum rises disproportionately more strongly, so that the
equivalent width is reduced by about a factor of two between quiescence and
the brightness maximum.
Similarly, for the two recent epochs that we have a measured brightness for,
the H$\alpha$ equivalent width increases by a factor of 1.6 
during a decline on a factor 0.48 in broadband brightness, showing again
that the continuum changes disproportionally more than the H$\alpha$
line flux.
}

\edit2{
The disproportionate rise of the continuum is even more
pronounced for the forbidden lines, 
}
\edit1{
the brighest of
which is [\ion{O}{1}] at 6300~{\AA}. From quiescence to the maximum,
a factor of $\approx$ 50 in broadband flux, the equivalent width
is reduced by a factor of $\approx$ 13. So while the line flux
in 
[\ion{O}{1}] 6300~{\AA} does increase somewhat during maximum, the broadband
flux increases 13 times more, showing that during maximum, the
conditions are less favorable for the emission of forbidden lines,
probably because of higher densities in the emitting regions.
}
\edit2{
Given the uncertainty of our assumption for the pre-outburst 
brightness when the 1993 Feb. 22 spectrum was taken, the
ESO-H$\alpha$ 99 spectral evolution is consistent with the
flux in those forbidden lines being unchanged during the 
outburst, and simply being diluted by increased continuum
or that line flux increasing by much less than the continuum.
}
\edit1{
Comparing just the 2019 March 17 (maximum) and 2019 April 05 (post maximum)
equivalent widths of [\ion{O}{1}]6300~{\AA} and the blended [\ion{S}{2}] lines, 
we clearly note that the equivalent width changes 
inversely proportional to the broadband flux, i.e., the absolute
line flux stays constant.
The accretion sensitive 
\citep{Muzerolle.1998.AJ.116.455M}
permitted CaII triplet lines show an increase of a
factor 1.3 during the decline by a factor of 0.48, meaning that the absolute
line flux has diminished, but not proportional to the continuum flux.
}

\edit2{
We are now comparing the spectral evolution of 
ESO-H$\alpha$~99 with several other EXor objects
and one higher luminosity eruptive variable.
}
\edit1{
The 
optical spectrum 
of ESO-H$\alpha$~99
and its changes during the outburst
are quite different from those observed in the 2008 outburst of the
prototypical EX Lupi.
\citet{ Kospal.2008.IBVS.5819.1K,
SiciliaAguilar.2012AA.544A.93S.EXLupi,
SiciliaAguilar.2015AA.580A.82S.EXLupi}
observed a 
}
\edit2{
very rich emission line spectrum, compared to other T Tauri stars, and
}
a substantial increase in the
number of metal emission lines during outburst, more than most other EXor-type outbursts, but note the absence of
forbidden lines. 
\edit2{
As a caveat to the latter statement, in high resolution spectra, emission of [\ion{O}{1}] was
detected in EX Lupi by \citet{2019ApJ...870...76B}
both in quiescence and, more strongly, during the 2008 outburst.
}

Similary, \citet{Holoien.2014.ApJ.785L.35H.ASASSN13db}
observed in the case of the 2013 EXor eruption of ASASSN13db that 
``during the outburst, the spectra are dominated by a forest of emission lines, mostly neutral metallic lines from \ion{Fe}{1}''.
The longer duration 2014-2017 outburst of this object was studied by 
\citet{SiciliaAguilar.2017.AA.607A.127S.ASASSN13db}
and they note again the similarity of the outburst spectrum to that of EX Lupi.

\edit2{
The spectral changes in these two objects are clearly different 
}
\edit1{
from the eruption of ESO-H$\alpha$ 99
where we observe a reduction in the prominence of the \ion{Fe}{1} line forest,
and a substantial reduction in the equivalent width, i.e. the importance
relative to the continuum, for the 
forbidden lines of [\ion{O}{1}] and [\ion{S}{2}].
}

\edit2{
In contrast, 
}
\edit1{
the evolution of the spectral lines during the
outburst of ESO-H$\alpha$ 99
has more similarities
to the higher luminosity case
PV Cep. 
\citet{Kun.2011.MNRAS.413.2689K}
found nearly constant flux, i.e. strongly increasing equivalent
widths, in forbidden lines during the fading.
However, they also
noted little change in the 
equivalent width of H$\alpha$ in the fading phase 
different from what we observe in ESO-H$\alpha$ 99.
}

\edit2{
We conclude that during the ESO-H$\alpha$ 99
outburst, its spectrum has changed substantially.
The H$\alpha$ line changes less than
the continuum during the outburst, and the forbidden [\ion{O}{1}] and [\ion{S}{2}]
lines change substantially less that the continuum, and are, in fact,
consistent with these forbidden line fluxes being constant}.

\edit2{
We conclude that
in ESO-H$\alpha$ 99,
the region emitting the forbidden lines may not be
affected at all by the processes leading to the continuum outburst. 
The increase in the optical and near-infrared continuum is qualitatively similar to the appearance of a strong 
continuum
during FUor outbursts. 
We speculate that in
in ESO-H$\alpha$ 99, the rise in the continuum is less than in typical FUor outbursts,
leaving the most prominent emission lines observable.
}

\section{Summary and Conclusions}

The optical (ATLAS $O$) light curve of ESO-H$\alpha$~99 shows a rise of about 4.4 magnitudes from the
pre-outburst average.
We also have indications for a previous small maximum of 2.5 mag amplitude
in 2016, and for a brief dip in brightness by 1.3 mag just prior
to the rise to the present maximum.
The high-cadence TESS light curve during a minor interim dip
in brightness just prior to reaching maximum light shows fluctuations
of order 10\% in flux and typical durations of a few
days, but without clear periodicity.

ESO-H$\alpha$~99 is associated with an optical and near-infrared
reflection nebula. There is one knot of line emission (MHO 1520) found
in the H$_2~1-0~S(1)$ emission line, but also indicated
in the optical R band. This is most likely a Herbig-Haro
object associated with collimated outflow activity.
\edit1{
The H$_2$ 1--0 S(1) emission line is also seen at the
position of the star.
}

\edit2{
ESO-H$\alpha$~99 shares the NIR colors and
mid-to-far IR SED of the most deeply embedded
EXors.
ESO-H$\alpha$~99 is a YSO of fairly high quiescent luminosity
(34 {$L_\odot$}), much higher than typical EXors,
and may be an intermediate mass star.
}

Two spectra during the present outburst shows many emission lines, in
particular H$\alpha$, \ion{Ca}{2}, and CO bandhead emission, 
making ESO-H$\alpha$~99 spectroscopically \edit1{similar} to other 
\edit1{deeply embedded} EXor outburst.
Comparison with a pre-outburst spectrum from 1993 shows,
however, that several emission lines that
had been present during quiescence are
partly diluted by continuum emission.
The H$\alpha$ equivalent width is largest in quiescence,
smallest near the light curve maximum and intermediate a few
weeks after the maximum.
The rise in overall brightness during this EXor event is largely
due to the disproportionate rise of the continuum compared to the
emission lines.
\edit1{
This effect is particularly strong for the forbidden [\ion{O}{1}]
and [\ion{S}{2}] lines.}
If the light curve does indeed show the typical evolution
of an EXor, the next observing season in late 2019 may show the
decline of brightness back to the pre-outburst level.

\acknowledgments
This work has made use of data from the European Space Agency (ESA) mission
{\it Gaia}
\footnote[3]{\url{https://www.cosmos.esa.int/gaia}}
and processed by the {\it Gaia}
Data Processing and Analysis Consortium (DPAC,
\footnote[4]{\url{https://www.cosmos.esa.int/web/gaia/dpac/consortium}}. Funding for the DPAC
has been provided by national institutions, in particular the institutions
participating in the {\it Gaia} Multilateral Agreement.
ATLAS observations and this work were supported by NASA grant NN12AR55G. The AAVSO Photometric All-Sky
Survey (APASS) was funded by the Robert Martin Ayers Sciences Fund. 
Infrared photometric data on ESO-H$\alpha$~99 were obtained at the IRIS telescope of the
Universit\"atssternwarte Bochum on Cerro Armazones,
which is operated under a cooperative agreement between
the "Astronomisches Institut, Ruhr Universit\"at Bochum", Germany
and the Institute for Astronomy, University of Hawaii, USA.
Construction of the IRIS infrared
camera was supported by the National Science Foundation under grant AST07-04954.
This work makes use of observations from the LCOGT network.
\edit1{
This paper uses data collected under the ESO/RUB – USB agreement at the Paranal Observatory.
}
This paper includes data collected by the TESS mission. Funding for the TESS mission is provided by the NASA Explorer Program.
This work is based in part on archival data obtained with the Spitzer Space Telescope, 
which is operated by the Jet Propulsion Laboratory, California Institute of Technology under a contract with NASA,
and on archival data from AKARI, a JAXA project with the participation of ESA.
The infrared spectrum was obtained with the SPEX instrument at the Infrared Telescope Facility, 
which is operated by the University of Hawaii under contract NNH14CK55B with the National Aeronautics and Space Administration.
Near-infrared imaging data from the WFCAM at the UKIRT observatory operated by the University of Hawaii were used in this paper.
PJV is supported by the National Science Foundation Graduate Research Fellowship Program Under Grant No. DGE-1343012.
\edit1{
CSK is supported by NSF grants AST-1515876, AST-1515927 and AST-181440.
This work made use of the ADS, Simbad, and VieziR.
We wish to thank the referee for constructive comments that helped improve this paper.
}

\vspace{5mm}
\facilities{ATLAS, FTN, Gaia, OCA:IRIS, IRTF, TESS, UKIRT}




\begin{thebibliography}{}

\bibitem[Abrahamyan et al.(2015)]{Abrahamyan.2015.AC.10.99A} Abrahamyan, H.~V., Mickaelian, A.~M., \& Knyazyan, A.~V.\ 2015, Astronomy and Computing, 10, 99

\bibitem[Aspin(2011a)]{Aspin.2011.AJ.141..196A} Aspin, C.\ 2011a, \aj, 141, 196 

\bibitem[Aspin(2011b)]{Aspin.2011.AJ.142.135A} Aspin, C.\ 2011b, \aj, 142, 135

\bibitem[Aspin et al.(2006)]{Aspin.2006.AJ.132.1298A} Aspin, C., Barbieri, C., Boschi, F., et al.\ 2006, \aj, 132, 1298 

\bibitem[Aspin et al.(2008)]{Aspin.2008.AJ.135.423A} Aspin, C., Beck, T.~L., \& Reipurth, B.\ 2008, \aj, 135, 423

\bibitem[Aspin, \& Reipurth(2009)]{Aspin.2009.AJ.138.1137A} Aspin, C., \& Reipurth, B.\ 2009, \aj, 138, 1137

\bibitem[Aspin, \& Sandell(1994)]{Aspin.1994.A&A.288.803A} Aspin, C., \& Sandell, G.\ 1994, \aap, 288, 803

\bibitem[Aspin et al.(2009)]{Aspin.2009.ApJ.692L.67A} Aspin, C., Reipurth, B., Beck, T.~L., et al.\ 2009, \apj, 692, L67

\bibitem[Aspin et al.(2009)]{Aspin.2009.AJ.137.2968A} Aspin, C., Greene, T.~P., \& Reipurth, B.\ 2009, \aj, 137, 2968

\bibitem[Aspin et al.(2010)]{2010ApJ...719L..50A} Aspin, C., Reipurth, B., Herczeg, G.~J., \& Capak, P.\ 2010, \apjl, 719, L50 

\bibitem[Audard et al.(2014)]{Audard.2014.prpl.conf.387A} Audard, M., {\'A}brah{\'a}m, P., Dunham, M.~M., et al.\ 2014, Protostars and Planets VI, 387

\bibitem[Banzatti et al.(2019)]{2019ApJ...870...76B} Banzatti, A., Pascucci, I., Edwards, S., et al.\ 2019, \apj, 870, 76

\bibitem[Casali et al.(2007)]{Casali2007} Casali, M., Adamson, A., Alves de Oliveira, C. et al. 2007, \aap, 467, 777

\bibitem[Cody et al.(2014)]{Cody.2014.AJ.147.82C} Cody, A.~M., Stauffer, J., Baglin, A., et al.\ 2014, \aj, 147, 82

\bibitem[Cody et al.(2017)]{Cody.2017.ApJ.836.41C} Cody, A.~M., Hillenbrand, L.~A., David, T.~J., et al.\ 2017, \apj, 836, 41

\bibitem[Contreras Pe{\~n}a et al.(2017b)]{ContrerasPena.2017.MNRAS.465.3039C} Contreras Pe{\~n}a, C., Lucas, P.~W., Kurtev, R., et al.\ 2017, \mnras, 465, 3039

\bibitem[Contreras Pe{\~n}a et al.(2017a)]{ContrerasPena.2017.MNRAS.465.3011C} Contreras Pe{\~n}a, C., Lucas, P.~W., Minniti, D., et al.\ 2017, \mnras, 465, 3011

\bibitem[Dame et al. (1987)]{Dame.1987.ApJ.322.706} Dame, T.M., Ungerechts, H., Cohen, R.S. et al. 1987, ApJ, 322, 706

\bibitem[Davis et al.(2010)]{Davis..2010A&A...511A..24D} Davis, C.~J., Gell, R., Khanzadyan, T., Smith, M.~D., \& Jenness, T.\ 2010, \aap, 511, A24

\bibitem[Eisloeffel et al.(1990)]{Eisloeffel.1990.A&A.237.369E} Eisloeffel, J., Solf, J., \& Boehm, K.~H.\ 1990, \aap, 237, 369

\bibitem[Eisloeffel et al.(1991)]{Eisloeffel.1991.ApJ.383L.19E} Eisloeffel, J., Guenther, E., Hessman, F.~V., et al.\ 1991, \apj, 383, L19

\bibitem[Epchtein et al.(1997)]{Epchtein.1997.Msngr.87.27E} Epchtein, N., de Batz, B., Capoani, L., et al.\ 1997, The Messenger, 87, 27

\bibitem[Fausnaugh et al.(2019)]{Fausnaugh.2019.arXiv190402171F} Fausnaugh, M.~M., Vallely, P.~J., Kochanek, C.~S., et al.\ 2019, arXiv e-prints, arXiv:1904.02171

\bibitem[Fischer et al.(2012)]{Fischer.2012.ApJ.756.99F} Fischer, W.~J., Megeath, S.~T., Tobin, J.~J., et al.\ 2012, \apj, 756, 99

\bibitem[Gaia Collaboration (2016)]{Gaia-2016A&A...595A...1G} Gaia Collaboration, Prusti, T., de Bruijne, J.~H.~J., et al.\ 2016, \aap, 595, A1 

\bibitem[Gaia Collaboration et al.(2018)]{Gaia.2018.A&A.616A.1G} Gaia Collaboration, Brown, A.~G.~A., Vallenari, A., et al.\ 2018, \aap, 616, A1

\bibitem[Gaia Collaboration et al.(2016)]{Gaia.2016.A&A.595A.2G} Gaia Collaboration, Brown, A.~G.~A., Vallenari, A., et al.\ 2016, \aap, 595, A2

\bibitem[Giannini et al. (2007)]{Giannini.2007.ApJ.671.470} Giannini, T., Lorenzetti, D., De Luca, M. et al. 2007, ApJ, 671, 470

\bibitem[Giannini et al.(2017)]{Giannini.2017.ApJ.839.112} Giannini, T., Antoniucci, S., Lorenzetti, D., et al.\ 2017, \apj, 839, 112

\bibitem[Greene et al.(1994)]{Greene.1994ApJ...434..614G} Greene, T.~P., Wilking, B.~A., Andre, P., et al.\ 1994, \apj, 434, 614

\bibitem[Hartmann et al.(2016)]{Hartmann.2016.ARA&A.54.135H} Hartmann, L., Herczeg, G., \& Calvet, N.\ 2016, \araa, 54, 135

\bibitem[Herbig(1962)]{Herbig.1962.AdA&A.1.47H} Herbig, G.~H.\ 1962, Advances in Astronomy and Astrophysics, 1, 47

\bibitem[Herbig(1977)]{Herbig.1977.ApJ.217.693H} Herbig, G.~H.\ 1977, \apj, 217, 693

\bibitem[Herbig(2007)]{Herbig.2007.AJ.133.2679H} Herbig, G.~H.\ 2007, \aj, 133, 2679

\bibitem[Herbig(2008)]{Herbig.2008.AJ.135.637H} Herbig, G.~H.\ 2008, \aj, 135, 637

\bibitem[Herbig et al.(2001)]{Herbig.2001.PASP.113.1547H} Herbig, G.~H., Aspin, C., Gilmore, A.~C., Imhoff, C.~L., \& Jones, A.~F.\ 2001, \pasp, 113, 1547 

\bibitem[Hillenbrand et al.(2019)]{Hillenbrand.2019.ApJ.874.82H} Hillenbrand, L.~A., Miller, A.~A., Carpenter, J.~M., et al.\ 2019, \apj, 874, 82

\bibitem[Hillenbrand et al.(1992)]{Hillenbrand.1992.ApJ.397.613H} Hillenbrand, L.~A., Strom, S.~E., Vrba, F.~J., et al.\ 1992, \apj, 397, 613

\bibitem[Hodapp \& Chini(2014)]{Hodapp.2014.ApJ.794.169H} Hodapp, K.~W., \& Chini, R.\ 2014, \apj, 794, 169

\bibitem[Hodapp et al.(2012)]{2012ApJ...744...56H} Hodapp, K.~W., Chini, R., Watermann, R., \& Lemke, R.\ 2012, \apj, 744, 56 

\bibitem[Hodapp et al.(1996)]{1996ApJ...468..861H} Hodapp, K.-W., Hora, J.~L., Rayner, J.~T., Pickles, A.~J., \& Ladd, E.~F.\ 1996, \apj, 468, 861 

\bibitem[Hodapp et al.(2010)]{Hodapp2010}
Hodapp, K. W., Chini, R., Reipurth, B., Murphy, M., Lemke, R.,
Watermann, R., Jacobson, S., Bischoff, K., Chonis, T., Dement, K.,
Terrien, R., \& Provence, S. 2010, Proc. SPIE 7735-45.

\bibitem[Hodgkin et al.(2009)]{Hodgkin.2009.MNRAS.394.675H} Hodgkin, S.~T., Irwin, M.~J., Hewett, P.~C., et al.\ 2009, \mnras, 394, 675

\bibitem[Holoien et al.(2014)]{Holoien.2014.ApJ.785L.35H.ASASSN13db} Holoien, T.~W.-S., Prieto, J.~L., Stanek, K.~Z., et al.\ 2014, \apjl, 785, L35

\bibitem[Holoien et al.(2019)]{Holoien.2019.arXiv190409293H} Holoien, T.~W.-S., Vallely, P.~J., Auchettl, K., et al.\ 2019, arXiv e-prints, arXiv:1904.09293

\bibitem[Kospal et al.(2008)]{Kospal.2008.IBVS.5819.1K} Kospal, A., Nemeth, P., Abraham, P., et al.\ 2008, Information Bulletin on Variable Stars, 5819, 1

\bibitem[Kun et al.(2011)]{Kun.2011.MNRAS.413.2689K} Kun, M., Szegedi-Elek, E., Mo{\'o}r, A., et al.\ 2011, \mnras, 413, 2689

\bibitem[Lasker et al.(2008)]{Lasker.2008.AJ.136.735L} Lasker, B.~M., Lattanzi, M.~G., McLean, B.~J., et al.\ 2008, \aj, 136, 735

\bibitem[Lehmann et al.(1995)]{Lehmann.1995.A&A.300L.9L} Lehmann, T., Reipurth, B., \& Brandner, W.\ 1995, \aap, 300, L9

\bibitem[Liseau et al. (1992)]{Liseau.1992.AA.265.577} Liseau, R., Lorenzetti, D., Nisini, B. et al. 1992, A\&A, 265, 577

\bibitem[Lomb(1976)]{Lomb1976}
Lomb, N. R. 1976, Ap\&SS, 39, 447

\bibitem[Lorenzetti et al.(2012)]{Lorenzetti.2012.ApJ.749.188} Lorenzetti, D., Antoniucci, S., Giannini, T., et al.\ 2012, \apj, 749, 188

\bibitem[Lorenzetti et al.(2015)]{Lorenzetti.2015.ATel.7935.1L} Lorenzetti, D., Giannini, T., Antoniucci, S., et al.\ 2015, The Astronomer's Telegram, 7935, 1

\bibitem[Lorenzetti et al.(2011)]{Lorenzetti.2011.ApJ.732.69L} Lorenzetti, D., Giannini, T., Larionov, V.~M., et al.\ 2011, \apj, 732, 69

\bibitem[Massi et al. (2007)]{Massi.2007.AA.466.1013} Massi, F., De Luca, M., Elia, D. et al. 2007, A\&A, 466, 1013

\bibitem[May, Murphy \& Thaddeus (1988)]{May.1988.AAS.73.51} May, J., Murphy, D.C., Thaddeus, P. 1988, A\&A Suppl. 73, 51

\bibitem[McLaughlin(1946)]{McLaughlin.1946.AJ.52.109M} McLaughlin, D.~B.\ 1946, \aj, 52, 109

\bibitem[Monet et al. (1998)]{Monet.1998.USNOA2.0} Monet, D. G. et al. (Flagstaff: US Nav. Obs.)

\bibitem[Munari et al.(2014)]{Munari.APASS.2014AJ....148...81M} Munari, U., Henden, A., Frigo, A., et al.\ 2014, \aj, 148, 81 

\bibitem[Murphy \& May (1991)]{Murphy.1991.AA.247.202} Murphy, D.C. \& May, J. 1991, \aap, 247, 202

\bibitem[Muzerolle et al.(1998)]{Muzerolle.1998.AJ.116.455M} Muzerolle, J., Hartmann, L., \& Calvet, N.\ 1998, \aj, 116, 455

\bibitem[Pettersson (2008)]{Pettersson.2008} Pettersson, B. 2008, in {\em Handbook of Star Forming Regions}, Vol. II, ed. Bo Reipurth, ASP, San Francisco, p. 43

\bibitem[Pettersson \& Reipurth (1994)]{Pettersson.1994.AAS.104.233} Pettersson, B. \& Reipurth, B. 1994, A\&A Suppl., 104, 233

\bibitem[Prizinzano et al. (2018)]{Prizinzano.2018.AA.617.A63} Prizinzano, L., Damian, F., Guarcello, M.G. et al. 2018, A\&A, 617, A63 

\bibitem[Ramolla et al.(2013)]{Ramolla.2013.AN.334.1115R} Ramolla, M., Drass, H., Lemke, R., et al.\ 2013, AN, 334, 1115

\bibitem[Rayner et al.(2003)]{Rayner.2003.PASP.115.362} Rayner, J.~T., Toomey, D.~W., Onaka, P.~M., et al.\ 2003, \pasp, 115, 362

\bibitem[Reipurth, \& Aspin(2004)]{Reipurth.2004.ApJ.606L.119R} Reipurth, B., \& Aspin, C.\ 2004, \apj, 606, L119

\bibitem[Reipurth \& Heathcote(1991)]{Reipurth.1991.AA.246.511} Reipurth, B., \& Heathcote, S.\ 1991, \aap, 246, 511

\bibitem[Ricker et al.(2015)]{Ricker.2015.JATIS.1a4003R} Ricker, G.~R., Winn, J.~N., Vanderspek, R., et al.\ 2015, Journal of Astronomical Telescopes, Instruments, and Systems, 1, 14003

\bibitem[Rieke, \& Lebofsky(1985)]{Rieke.1985.ApJ.288.618R} Rieke, G.~H., \& Lebofsky, M.~J.\ 1985, \apj, 288, 618

\bibitem[Rodgers, Campbell, \& Whiteoak (1960)]{Rodgers.1960.MNRAS.121.103} Rodgers,A.W., Campbell, C.T., Whiteoak, J.B. 1960, MNRAS, 121, 103 

\bibitem[Sandqvist \& Lindroos (1976)]{Sandqvist.1976.AA.53.179} Sandqvist, Aa. \& Lindroos, K.P. 1976, A\&A, 53, 179

\bibitem[Scargle(1982)]{Scargle1982}
Scargle, J. D. 1982, \apj, 263, 835

\bibitem[Sicilia-Aguilar et al.(2015)]{SiciliaAguilar.2015AA.580A.82S.EXLupi} Sicilia-Aguilar, A., Fang, M., Roccatagliata, V., et al.\ 2015, \aap, 580, A82

\bibitem[Sicilia-Aguilar et al.(2012)]{SiciliaAguilar.2012AA.544A.93S.EXLupi} Sicilia-Aguilar, A., K{\'o}sp{\'a}l, {\'A}., Setiawan, J., et al.\ 2012, \aap, 544, A93

\bibitem[Sicilia-Aguilar et al.(2017)]{SiciliaAguilar.2017.AA.607A.127S.ASASSN13db} Sicilia-Aguilar, A., Oprandi, A., Froebrich, D., et al.\ 2017, \aap, 607, A127

\bibitem[Siwak et al.(2013)]{Siwak.2013.MNRAS.432.194S.FU.Ori} Siwak, M., Rucinski, S.~M., Matthews, J.~M., et al.\ 2013, \mnras, 432, 194

\bibitem[Siwak et al.(2018)]{Siwak.2018.AA.618A.79S.FU.Ori} Siwak, M., Winiarski, M., Og{\l}oza, W., et al.\ 2018, \aap, 618, A79

\bibitem[Skrutskie et al.(2006)]{Skrutskie2006} Skrutskie, M. F., Cutri, R. M., Stiening, et al. 2006, \aj, 131, 1163

\bibitem[Stauffer et al.(2014)]{Stauffer.2014.AJ.147.83S} Stauffer, J., Cody, A.~M., Baglin, A., et al.\ 2014, \aj, 147, 83

\bibitem[Strafella et al. (2010)]{Strafella.2010.ApJ.719.9} Strafella, F., Elia, D., Campeggio, L. et al. 2010, ApJ, 719, 9

\bibitem[Strafella et al. (2015)]{Strafella.2015.ApJ.798.A104} Strafella, F., Lorenzetti, D., Giannini, T. et al. 2015, ApJ, 798, A104

\bibitem[Strai{\v{z}}ys et al.(2008)]{Straizys.2008.BaltA.17.125} Strai{\v{z}}ys, V., Corbally, C.~J., \& Laugalys, V.\ 2008, Baltic Astronomy, 17, 125

\bibitem[Tokunaga, Simons, \& Vacca(2002)]{Tokunaga2002} Tokunaga, A. T., Simons, D. A., \& Vacca, W. D. 2002, \pasp, 114, 792

\bibitem[Tokunaga \& Vacca(2005)]{Tokunaga..MKO.filters..2005PASP..117..421T} Tokunaga, A.~T., \& Vacca, W.~D.\ 2005, \pasp, 117, 421

\bibitem[Tonry et al.(2018)]{Tonry.2018PASP..130f4505T} Tonry, J.~L., Denneau, L., Heinze, A.~N., et al.\ 2018, \pasp, 130, 064505 

\bibitem[Vallely et al.(2019)]{Vallely.2019.MNRAS.1385V} Vallely, P.~J., Fausnaugh, M., Jha, S.~W., et al.\ 2019, \mnras, 1385

\bibitem[Wegner(2014)]{Wegner.2014.AcA.64.261W} Wegner, W.\ 2014, \actaa, 64, 261

\bibitem[Yamaguchi et al. (1999)]{Yamaguchi.1999.APSJ.51.775} Yamaguchi, N., Mizuno, N., Saito, H. et al. 1999, PASJ, 51, 775


\end{thebibliography}
\end{document}